\documentclass[11pt]{report}
\usepackage{graphicx} 
\usepackage{subcaption}
\usepackage{pdfpages}
\usepackage{adjustbox}
\usepackage{hyperref}
\usepackage{titlesec}
\usepackage{fancyhdr}
\usepackage{float}
\usepackage{xcolor}
\usepackage{sectsty}
\usepackage[a4paper]{geometry}
\usepackage[myheadings]{fullpage}
\usepackage[font=footnotesize, labelfont=bf]{caption}

\usepackage[
    backend=biber,
    style=numeric,
    sorting=none
]{biblatex}
\title{Elevating Academic Administration: A Comprehensive Faculty Dashboard for Tracking Student Evaluations and Research}
\author{Musa Azeem}
\date{May 2024}

\definecolor{garnet}{RGB}{115,0,10}
\definecolor{rose}{RGB}{132,66,71}
\sectionfont{\color{garnet}}
\subsectionfont{\color{garnet}}

\let\clearpage\relax
\titleformat{\chapter}[display]{\color{rose}\Large\bfseries}{Part\ \thechapter}{0pt}{\Huge}
\titlespacing{\chapter}{0pt}{10pt}{15pt}

\begin{document}


\includepdf[pages=-]{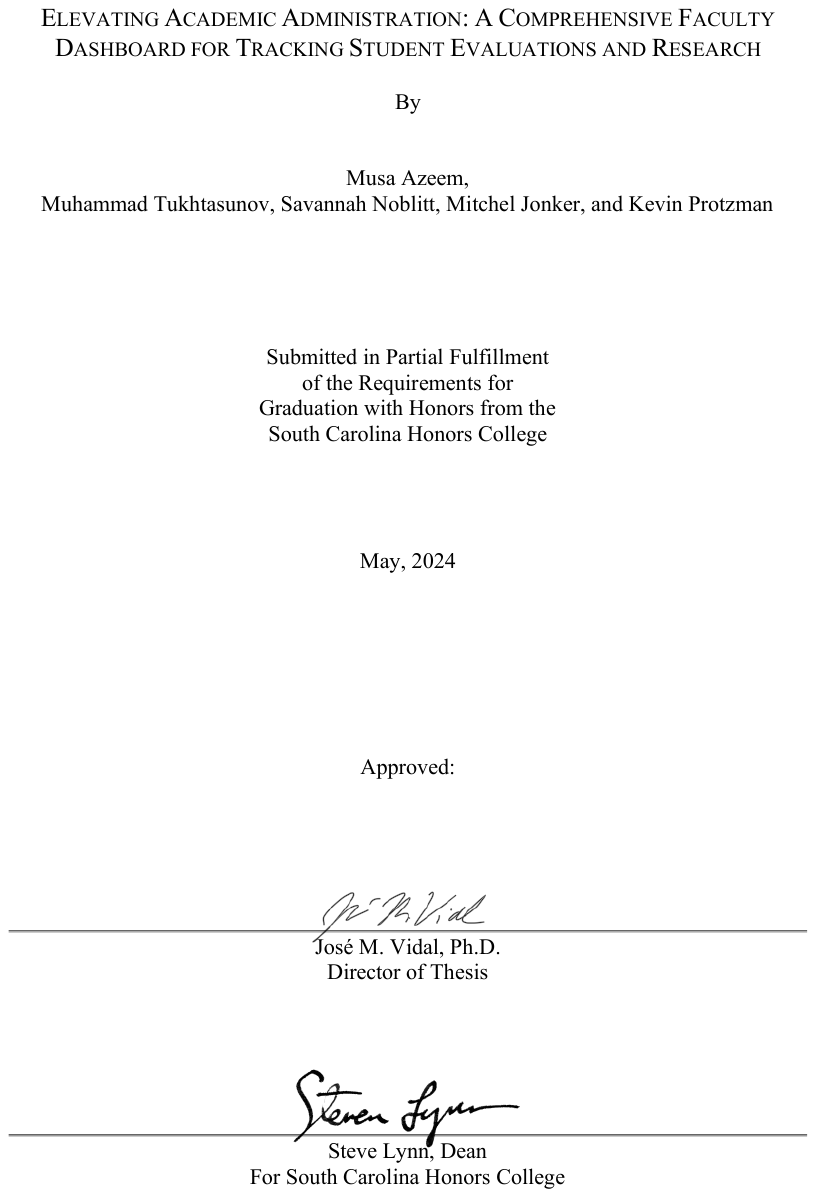}

\begin{abstract}
    The USC Faculty Dashboard is a web application designed to revolutionize how department heads, professors, and instructors monitor progress and make decisions, providing a centralized hub for efficient data storage and analysis. Currently, there's a gap in tools tailored for department heads to concisely manage the performance of their department, which our platform aims to fill. The USC Faculty Dashboard offers easy access to upload and view student evaluation and research information, empowering department heads to evaluate the performance of faculty members and seamlessly track their research grants, publications, and expenditures. Furthermore, professors and instructors gain personalized performance analysis tools, with full access to their own data as well as curated access to peer data to assess their relative performance. The source code as well as the link to the deployed application can be found at \url{https://github.com/SCCapstone/K3MS}.
\end{abstract}


\newpage
\chapter*{Contributions}
    \thispagestyle{empty}
    This project was developed as a Senior Capstone project at UofSC. Five students contributed to its development: Musa Azeem, Muhammad Tukhtasunov, Savannah Noblitt, Mitchel Jonker, and Kevin Protzman. Furthermore, Dr. Jos\'e M. Vidal contributed as the professor of the capstone course and advisor for this project, and Dr. Homayoun Valafar was involved as the initiator, primary stakeholder, and key end user of the platform. 

\newpage
\tableofcontents
\thispagestyle{empty}

\newpage
\setcounter{page}{1}
\chapter{The Platform \& What it Solves}
    \thispagestyle{fancy}
    
    At the heart of this project lies the ever-present and consistently pressing matter of data management. Specifically, it revolves around the intricate task of overseeing the course performances, research endeavors, and expenditures of professors and instructors. This responsibility is shouldered by not only the professors themselves, but, to a greater extent, the executive chairs of academic departments overseeing them. As of now, faculty members face numerous obstacles in collecting and analyzing this data to construct meaningful improvements. The development of a comprehensive, concise, and easy-to-use application to streamline this effort is essential to maintain the efficiency and effectiveness of faculty as academic departments continue to grow.

    \section{Academic Administration Now} \label{problem}
        Professors and academic administrators juggle many responsibilities, but here we discuss two intertwined data management challenges they face: various sources of data and the lack of an effective method to view and analyze collected information.
    
        \subsection{Various Sources of Data}
            The first challenge currently hindering the efficiency of professors and department chairs is the collection of data from faculty. There are four essential types of information collected by department chairs for each professor and instructor:
        
            \begin{itemize}
                \item \textbf{Student Evaluations:} The performance of each faculty member in the courses they teach
                \item \textbf{Publications:} New publications produced by a research group
                \item \textbf{Grants:} Money that a faculty member is producing in the form of awarded grants 
                \item \textbf{Expenditures:} Amount that a research group has spent in a given year
            \end{itemize}
        
            The first component of this obstacle is reporting this information, which frustratingly requires the use of various disconnected platforms. The Department of Computer Science \& Engineering at the University of South Carolina, for instance, mandates the use of USCeRA, Google Forms, and internal services to report publications, expenditures, and student evaluations. 
            
            The second, resultant, component is the collection of data from these various sources. This task falls upon department chairs, who must again visit each of these platforms to manually retrieve and organize data for future use.
        
        \subsection{Obstacles in Analyzing Data}
            The other major data management challenge currently faced by faculty of academic departments involves the subsequent step once data has been collected: viewing and analyzing data to determine progress, detect patterns, and construct meaningful plans for improvements going forward. This essential administrative step is hindered by the lack of tailored tools in analyzing this sort of data. Department chairs must develop their own methods to detect any patterns or compare faculty members to assess relative performance. Likewise, professors and instructors have no way to securely evaluate their performance relative to their peers, to see how they compare. Whether its a professor trying to improve themselves or an administrator striving to move their department in a positive direction, the efficient and effective analysis of collected data is a vital, but currently unsatisfactory, component of the academic environment. 
            
    \section{The Solution: An All-in-One Faculty Dashboard}
        There a prominent gap in tools curated to address the challenges outlined in $\S$\ref{problem}. No sufficient platform exists to facilitate the reporting, collection, and analysis of student evaluation and research information all in one place. Such a tool is essential to maintain and improve academic administration techniques, especially as departments continue to grow. The USC Faculty Dashboard aims to fill this gap with all necessary features to serve as a single, comprehensive, and streamlined web application for academic administration.
            
        \subsection{Centralized Platform for Uploading and Collecting Data}
\begin{figure}[h]
                \centering
                    \includegraphics[width=0.5\linewidth]{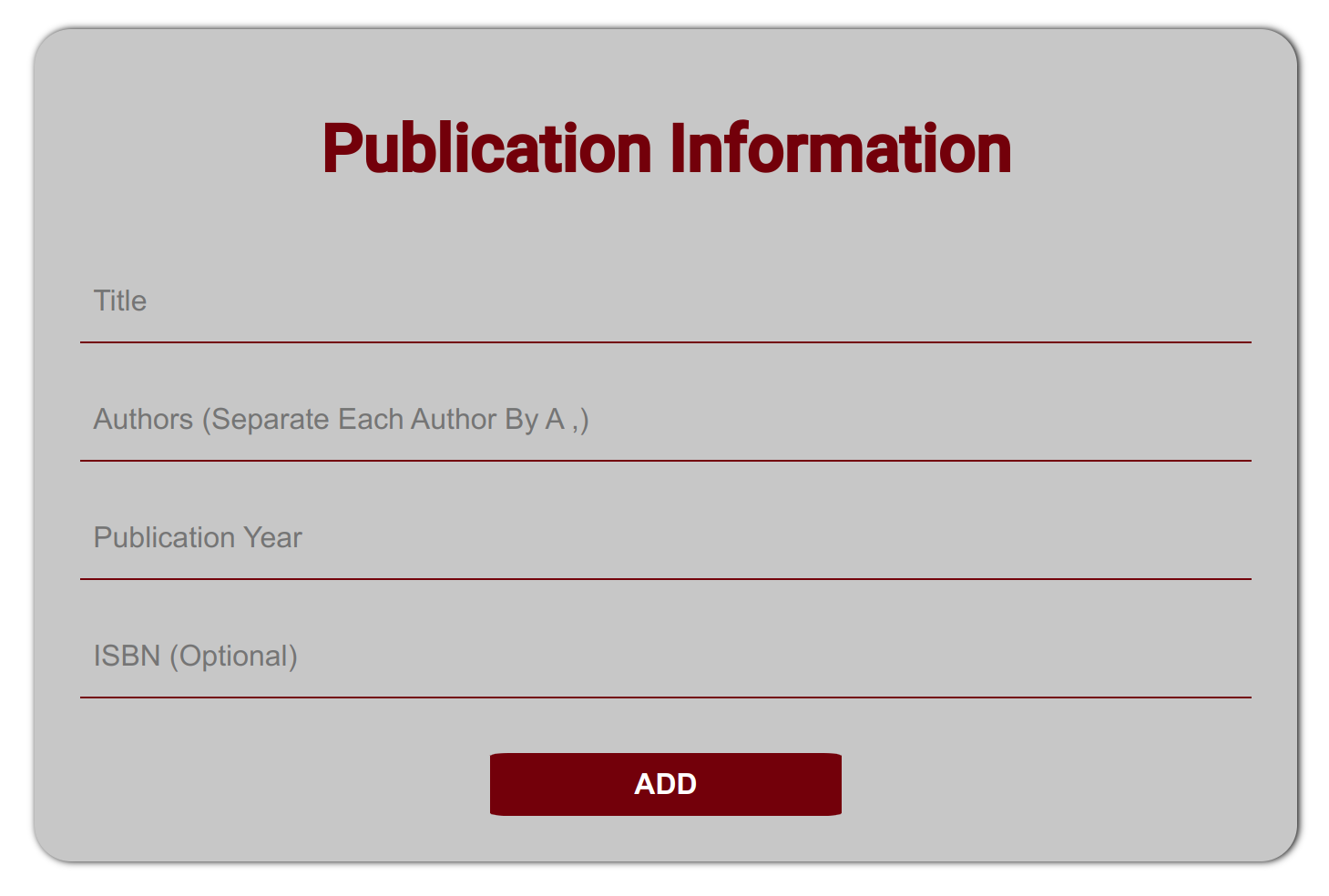}
                \caption{A form on our platform where users can report new publications, which are instantly made available to see for the user and their administrator.}
                \label{fig:pubupload_form}
            \end{figure}

            To maintain and improve the efficiency of instructors, professors, and department heads, our platform serves as a central location to upload student evaluations and report new publications, grants, and research expenditures. A streamlined and concise user interface (UI) replaces the numerous data reporting platforms currently required. To facilitate this, forms with consistent UI for each type of data are integrated into the platform, an example of which is shown in Figure \ref{fig:pubupload_form}.
            
             Concurrently, our platform also streamlines the process of collecting this information. Via the use of a centralized, integrated database for the platform, all data reported by faculty is instantly available for themselves and their department heads. As illustrated in Figures  \ref{fig:studentevals} and \ref{fig:researchinfo-grants}, all evaluation and research data is centralized for administrators to view, circumventing the need for collecting and organizing any information.
        
        \subsection{Enabling Efficient and Concise Analysis of Data}
\begin{figure}[h]
                \centering
                \includegraphics[width=0.75\linewidth]{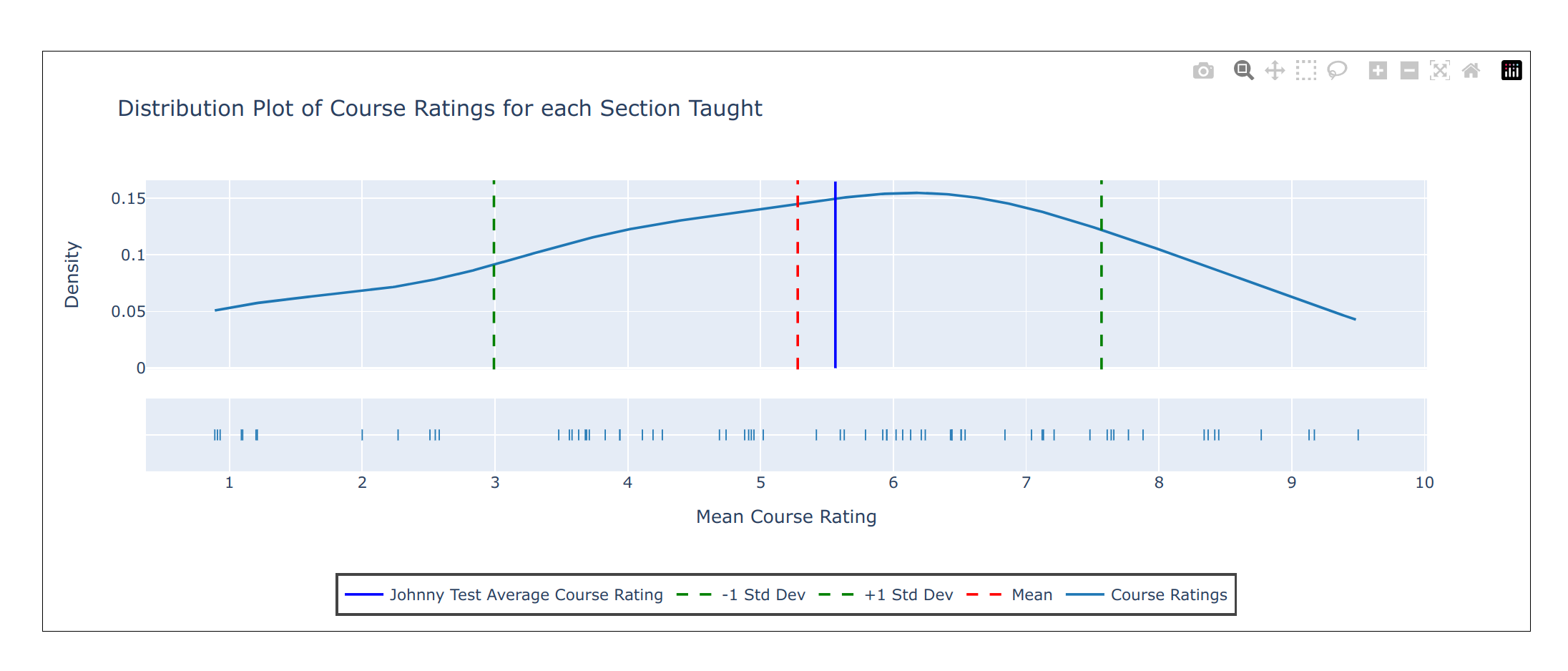}
                \caption{Kernel Density Estimation (KDE) distribution plot of the average course rating the instructor of every section of a certain course received. This plot is interactive and integrated in the Course Analytics page of our platform. Department chairs are given the option to see this plot for anybody in their department (choose who's average course rating is shown in blue). Other faculty can only see it for themselves.}
                \label{fig:dist_plot}
            \end{figure}

            Another core feature of The USC Faculty Dashboard is the facilitation of efficient, concise, and meaningful data analysis. Simply by aggregating all related data onto one platform, our app greatly improves department chairs' ability to efficiently review course performance and research endeavors of faculty members and quickly compare them to each other. To further facilitate meaningful data analysis, our platform performs statistical analyses of uploaded student evaluation data. Components on our platform (shown in Figure \ref{fig:team_assess}) include percentile calculations to enable department chairs to assess the relative performance of each faculty member across all courses they teach. Furthermore, distribution plots, as shown in Figure \ref{fig:dist_plot}, are available for both administrators and faculty members to visualize performance in direct comparison to all other instructors of a course. These plots are completely anonymized for non-administrators, such that faculty members are provided insight into their performance while preserving the privacy of other professors.

            See Appendix \ref{appendixA} for a full, in-depth description of all features and pages featured on our platform.

\chapter{Methods}
    \thispagestyle{fancy}
    This project was developed as a part of a Senior Capstone project by five students at the University of South Carolina. It was developed over the course of the Fall 2023 and Spring 2024 semesters. This section provides an overview of the collaborative methods, stages of development, use of feedback, and technologies utilized for this project.

    \section{Collaboration}
        This project was the collaborative effort of five students. This team met regularly once a week and liberally communicated as needed outside of these meetings. The development process followed the ``Agile'' methodology \autocite{fowler2001agile}. Throughout the development process, ``issues'' were defined as elements of the project queued for development or requiring attention (eg. ``develop a login page'' or ``fix this component that is not working as intended''). During weekly meetings, the team would define new issues, assign issues to team members, review completed issues, and discuss any roadblocks or upcoming milestones. Issues were assigned to members of the team based on internal discussion. Each issue would be resolved independently by one or more team members--with assistance from other members provided as needed--and progress for each issue was reviewed by all team members before incorporating changes into the main developing project. Throughout the entire development process, the Git version control software along with the GitHub development platform was utilized to collaborate and track progress \autocite{Git, GitHub}.

    \section{Stages of Development \& Timeline}
\begin{figure}[h]
            \centering
            \includegraphics[width=0.75\linewidth]{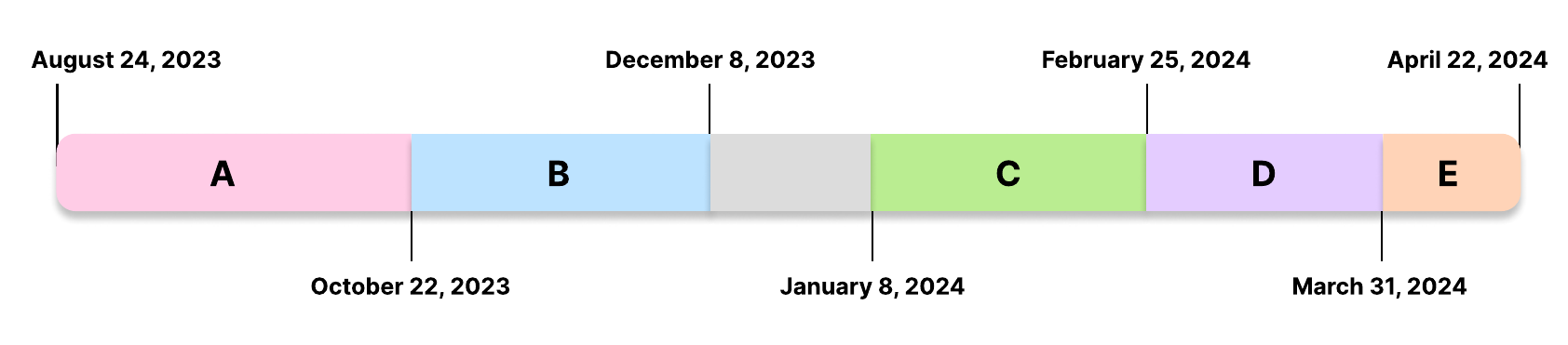}
            \caption{Timeline for the development stages of this project. The following stages correspond to labels in the figure: Planning and Preliminary Work (\textbf{A}), Proof of Concept (\textbf{B}), Beta Release (\textbf{C}), Release Candidate 1 (\textbf{D}), and 1.0 Release (\textbf{E}). The unlabeled region corresponds to winter break between semesters when no development work was done.}
            \label{fig:timeline}
        \end{figure}

        The development of this project was divided into five major milestones, which served to organize and maintain the pace of project development. An overview of the timeline for these stages is illustrated in Figure \ref{fig:timeline}.

        \subsection{Planning and Preliminary Work}\label{planning}
            The first development stage involved planning steps and preliminary work to organize future development. It was completed by October 22, 2023. 
            
            For this milestone, the team first brainstormed a list of pages to be included in the web application, as well as a high level description of the features to be available on each page. Next, the Figma software was utilized to develop prototype designs of each page of the final platform \autocite{figma}. These pages were used as references to guide the development process.
            
            This stage also involved formally defining all requirements of the platform. This meant defining an intricate list of all features and functionality to be incorporated into the final product. Features were classified as Required, Desired, or Aspirational, which determined the priority of their development (priority descending).
            
            Finally, during this stage, the team conducted research to determine which technologies would best suit the development of this project. The technologies incorporated in this project were chosen based on the characteristics of the platform being developed as well as the previous experiences of the team. Further details on these decisions and the technologies themselves are provided in $\S$\ref{tech}. Once technologies were chosen, the technical architecture of each component of our platform was laid out as the final step of this stage.
        
        \subsection{Proof of Concept}
            The next major milestone was the Proof of Concept, which was released on December 8, 2023. For this milestone, the team developed a prototype of the platform, including only the features necessary to establish to the client the viability of the final solution. For this release, the backbone of the application (components necessary for any web application) was implemented. Additionally, forms to upload research data and components to view them, as seen in Figures \ref{fig:researchinfo-grants} and \ref{fig:uploaddata}, were implemented along with the database functionality to store this data.
        
        \subsection{Beta Release}
            The Beta Release Milestone was released on February 25, 2024, and served as a Minimally Viable Product (MVP). This release consisted of all major features and was a sufficient, however lacking, final product. At this point, the client was given the opportunity to critique the project at its current state and provide feedback to steer the development in the desired direction. 
        
        \subsection{Release Candidate 1}\label{RC1}
            For the Release Candidate 1 milestone, all features of the app were implemented. It was released on March 31, 2024. This release served as a final release candidate provided to the client to critique and supply feedback preceding the final development stages. Additionally, here is when Quality Assurance (QA) testing was performed to rigorously ensure the platform behaved as expected.
        
        \subsection{1.0 Release}
            The final product was released on April 22, 2024. This milestone involved incorporating any requested changes and resolving any issues raised by QA.

    \section{Use of Feedback}
        Responding to feedback played a critical role in the development of this project. Throughout the development process, feedback was received from the client, Dr. Homayoun Valafar, the course instructor, Dr. Jos\'e M. Vidal, and, at one point, fellow students. Feedback from all sources was heavily relied upon to guide the development of the platform.
        
        \subsection{Working with a Client}
            From the initiation of this project's idea to the specification of the desired features, Dr. Valafar was involved as the primary stakeholder and motivating client for the USC Dashboard Platform. After initially brainstorming the requirements of the app, as described in $\S$\ref{planning}, feedback from Dr. Valafar was used to refine them until they met the desired criteria. Furthermore, feedback from the client was received at critical milestones to ensure the project was headed in the desired direction.
        
        \subsection{Course Instructor Feedback}
            The key advisor throughout the development of this project was Dr. Vidal, the professor for the Capstone course. After each major milestone, feedback was received from Dr. Vidal in the form of issues to be resolved. These issues were critical to track bugs and ensure our platform followed the best industry standards throughout the development process. Through the resolution of these issues as a part of our Agile methodology, Dr. Vidal's feedback was directly incorporated into the project.
        
        \subsection{Peer Quality Assurance Reviews}
            Finally, following the completion of the Release Candidate 1 Milestone ($\S$\ref{RC1}), we were provided the opportunity to receive QA testing from fellow students. These students rigorously tested our application to ensure it was free of bugs, was performing as expected, and was functional from the perspective of someone completely unfamiliar with the platform. Feedback from these students was received as issues and resolved as a part of our Agile process.

    \section{Technology}\label{tech}
        One of the key decisions for any software development project is the technology stack utilized in implementing the product. Our software consists of three primary components: the frontend, the backend, and the database. As briefly introduced in $\S$\ref{planning}, each component requires careful consideration of the technologies used for its implementation. This section lays out the technologies we chose and the contributing factors towards those decisions.
    
        \subsection{Frontend: JavaScript, React, \& Plotly}
            The frontend of our software is the component that users see and interact with. Virtually every frontend web application is built using the JavaScript programming language\footnote{Many applications are developed using alternative versions of JavaScript as well, such as TypeScript.} \autocite{javascript}. However, there are numerous frameworks built upon JavaScript to enable more rapid and advanced web application development, each with its own benefits and drawbacks. One such framework, React, is the most popular among JavaScript programmers and was chosen for this project \autocite{react, react_stats}. The decision to develop with React was based not only on our team's previous experience working with the framework, but also on React's component-based architecture that enables efficient and organized development and meets all requirements for an application such as ours.

            Another vital technology utilized in the frontend component of our software was Plotly \autocite{plotly}. Plotly is a multi-platform software library that enables the generation of interactive, customizable plots for various forms of data. Plotly's ability to integrate with both Python and JavaScript served our application's requirements perfectly and was utilized to create the distribution plots shown in Figure \ref{fig:dist_plot}.

        \subsection{Backend: Python \& Flask}
            The backend component of our software is never seen by the user, but handles all data processing, authentication, and interactions with the database. It also serves as a vital layer of security between the user and the sensitive data stored in our database, ensuring users only interact with data they have authority over. Options for backend technologies are much more extensive than frontend, with a virtually endless list of programming languages and frameworks used in backend systems. For our purposes, we chose to develop our backend in Python, using the Flask microframework \autocite{python, flask}. 

            Python was chosen due to its support of multiple frameworks--discussed next--and, primarily, its extensive library of data processing and analysis modules. Python modules, including Pandas, Numpy, and Scipy, were vital to the development of features such as Excel sheet processing and percentile calculations.

            Within Python, there exist multiple frameworks to facilitate the development of a backend server. Our team decided to develop our software with Flask, a microframework capable of serving all our application's needs. Flask's features, including login authentication and database abstraction, enabled efficient and modern development of our software.
        \subsection{Database: MySQL}
            The final core component of our software is the database. The database component is responsible for storing all uploaded data and supplying specific information in response to user activity. There are two primary categories of database technologies: Relational and Non-relational \autocite{jatana2012survey}. For our project, a relational database was chosen due to its data aggregation features and tabular organization that fit our data. Specifically, we implemented our database using MySQL, a fast and lightweight system very popular among web applications \autocite{mysql}.  
        
\chapter{Conclusions}
    \thispagestyle{fancy}

    \section{Implications}
        The development of the USC Faculty Dashboard marks the first stage of filling the current gap in academic administrative technology. Academic departments consistently strive to improve their impact on their field of research and their students. Through this platform, department chairs and faculty members can more efficiently manage their data and effectively extract meaningful insights into what the data means. After adopting our platform, we hope faculty members will begin to see improvements in their workflows and ability to improve themselves and their departments.

    \section{Future Work}
        Like all software applications, the requirements and use cases of this platform will continue to evolve as users begin to utilize the application in their regular workflows. A vital factor in the future development of this project will simply be maintaining the application to keep up with new feature requests and discovered bugs. 

        Another key component of the future development of this project will be its expansion for a larger user base. Currently, the platform is developed primarily for the Department of Computer Science \& Engineering at the University of South Carolina. In the future, it will be expanded to provide services to other departments here at USC and, eventually, to other universities.

 \newpage
\printbibliography[title={References}]

\newpage
\appendix
\pagenumbering{roman}
\fancyhf{}
\renewcommand{\headrulewidth}{0pt}
\pagestyle{fancy} 
\fancyhead{}
\fancyfoot[R]{\thepage}

\titleformat{\chapter}[display]{\color{rose}\Large\bfseries}{\chaptertitlename\ \thechapter}{0pt}{\Huge}

\chapter{Pages of the Webapp}\label{appendixA}
    \thispagestyle{fancy}
    \noindent This appendix contains screenshots of each major page in the application.

\section{Dashboard}
    \begin{figure}[H]
        \centering
        \begin{adjustbox}{frame=0.25pt}
            \includegraphics[width=1\linewidth]{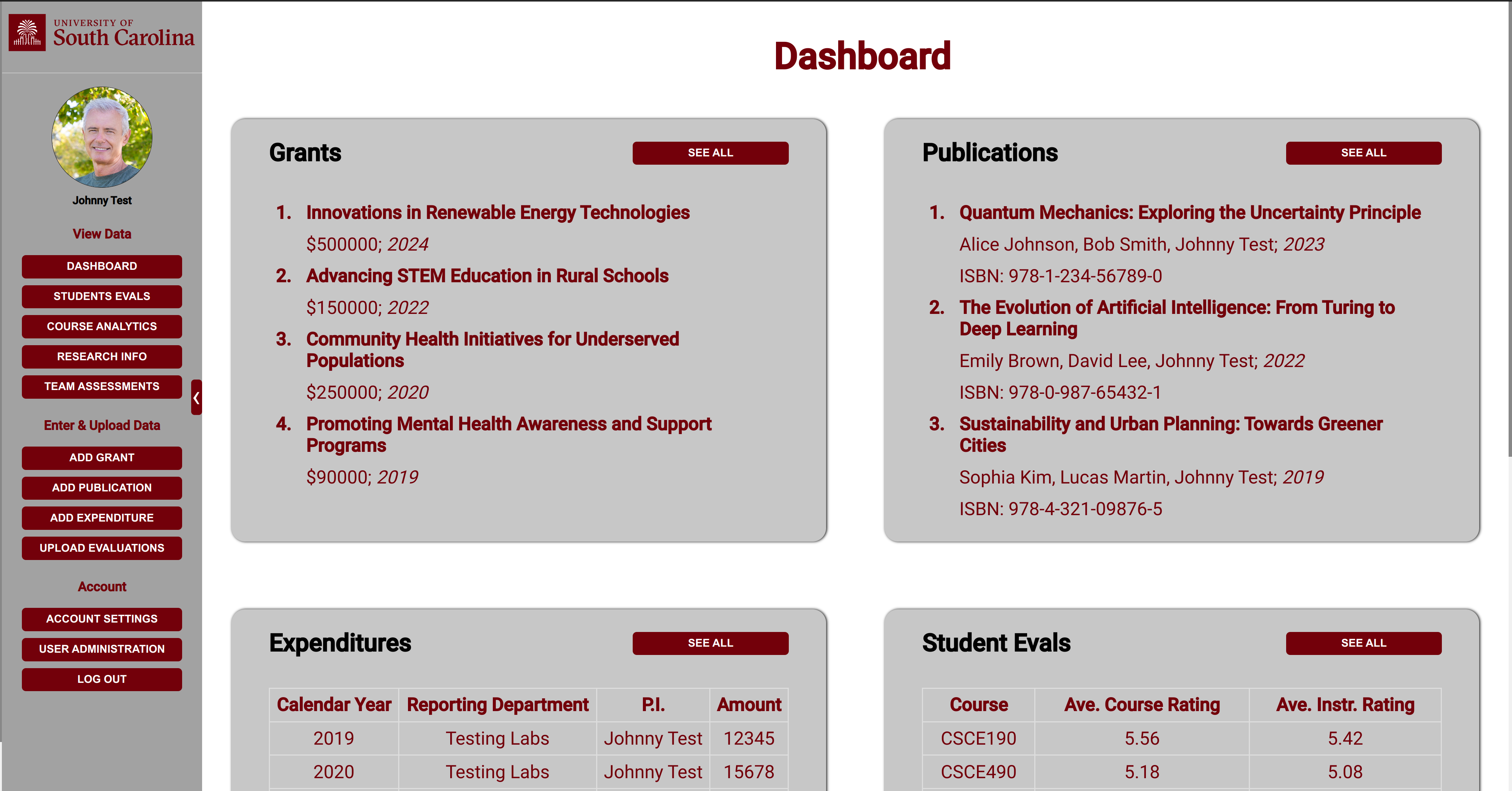}
        \end{adjustbox}
        \caption{Dashboard page of our platform. This page shows previews of each major page to provide information at a glance. Links to other pages are found in the collapsible navigation bar on the left side.}
        \label{fig:dashboard}
    \end{figure}

\section{Student Evalutions}
    \begin{figure}[H]
        \centering
        \begin{adjustbox}{frame=0.25pt}
            \includegraphics[width=1\linewidth]{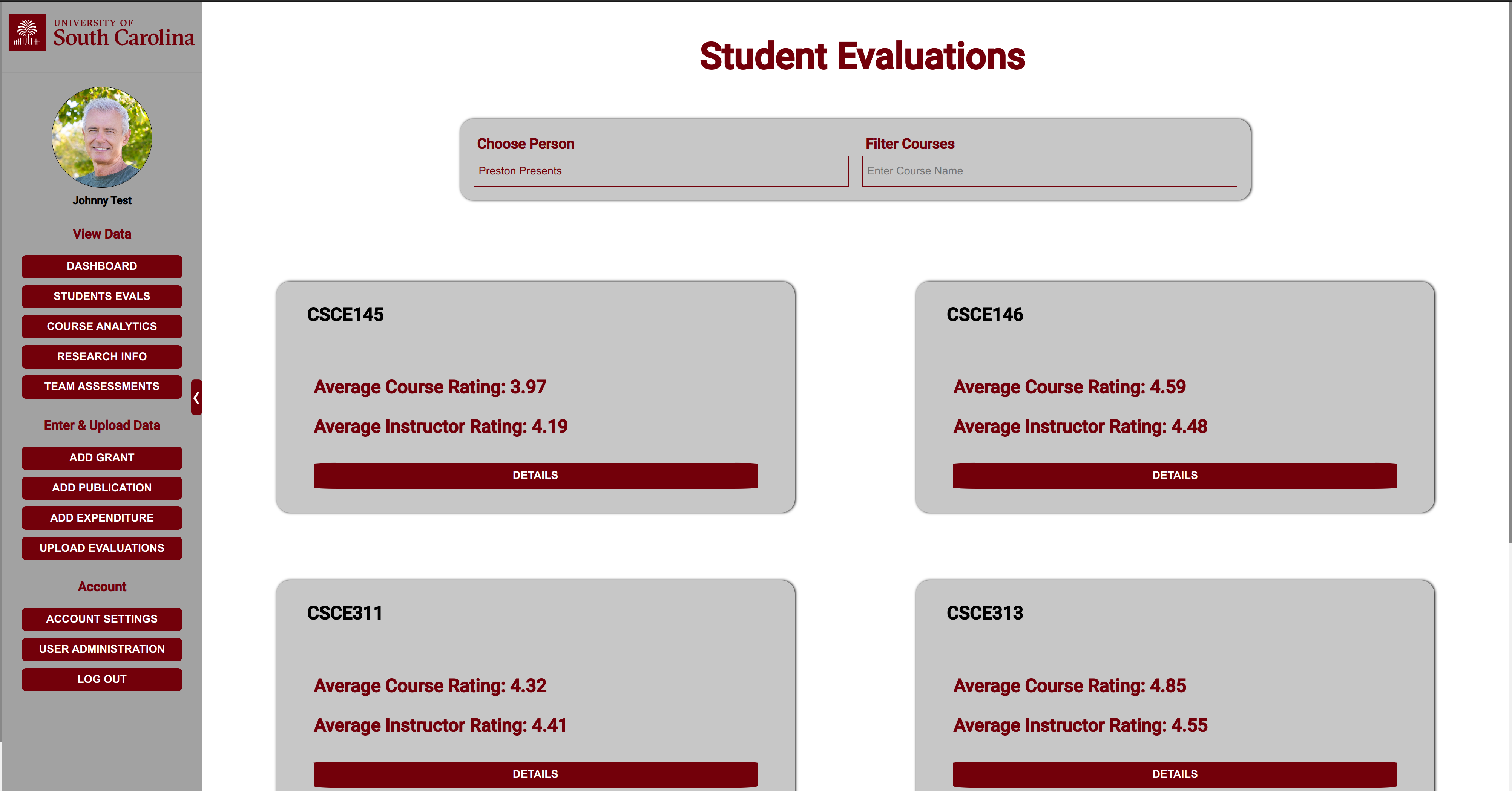}
        \end{adjustbox}
        \caption{The Student Evaluations page on our platform. This page shows the average course and instructor ratings received from students for each course a user teaches. As shown in the image, an administrator has the authority to view evaluations for themselves and members of their department (here, department chair Johnny Test views student evaluations for professor Preston Presents.}
        \label{fig:studentevals}
    \end{figure}

    \begin{figure}[H]
    \centering
        \begin{adjustbox}{frame=0.25pt}
            \includegraphics[width=1\linewidth]{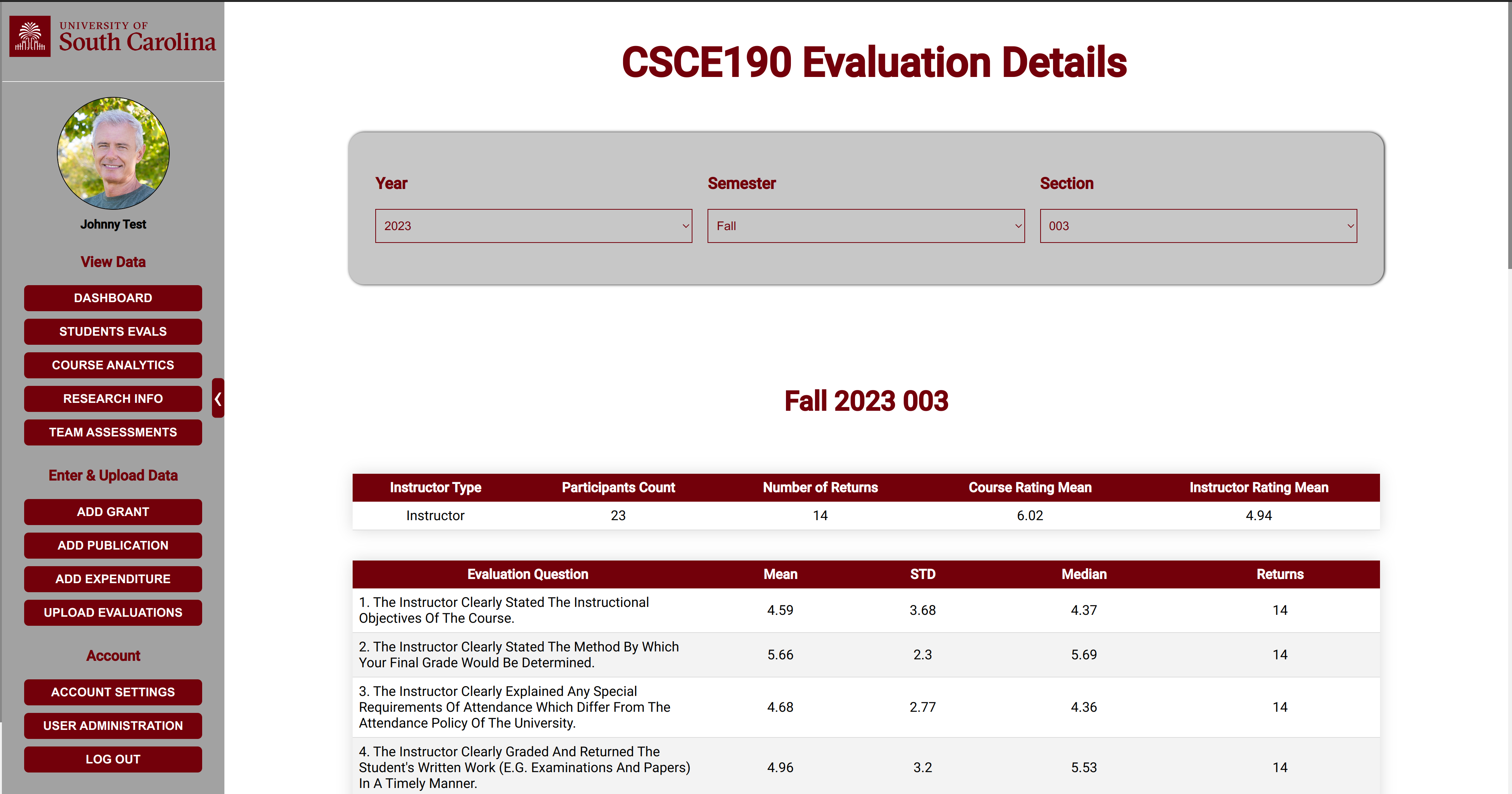}
        \end{adjustbox}
        \caption{The Student Evaluations Details page on our platform is available for each course a user teaches. This page is accessible through the buttons on cards in the Student Evaluations Page. Here, users can see the aggregated answers to each question provided to their students during evaluations. They are able to choose any section they taught the course for to see these details.}
        \label{fig:studentevaldetails}
    \end{figure}

\section{Course Analytics}
    \begin{figure}[H]
        \centering
        \begin{adjustbox}{frame=0.25pt}
            \includegraphics[width=1\linewidth]{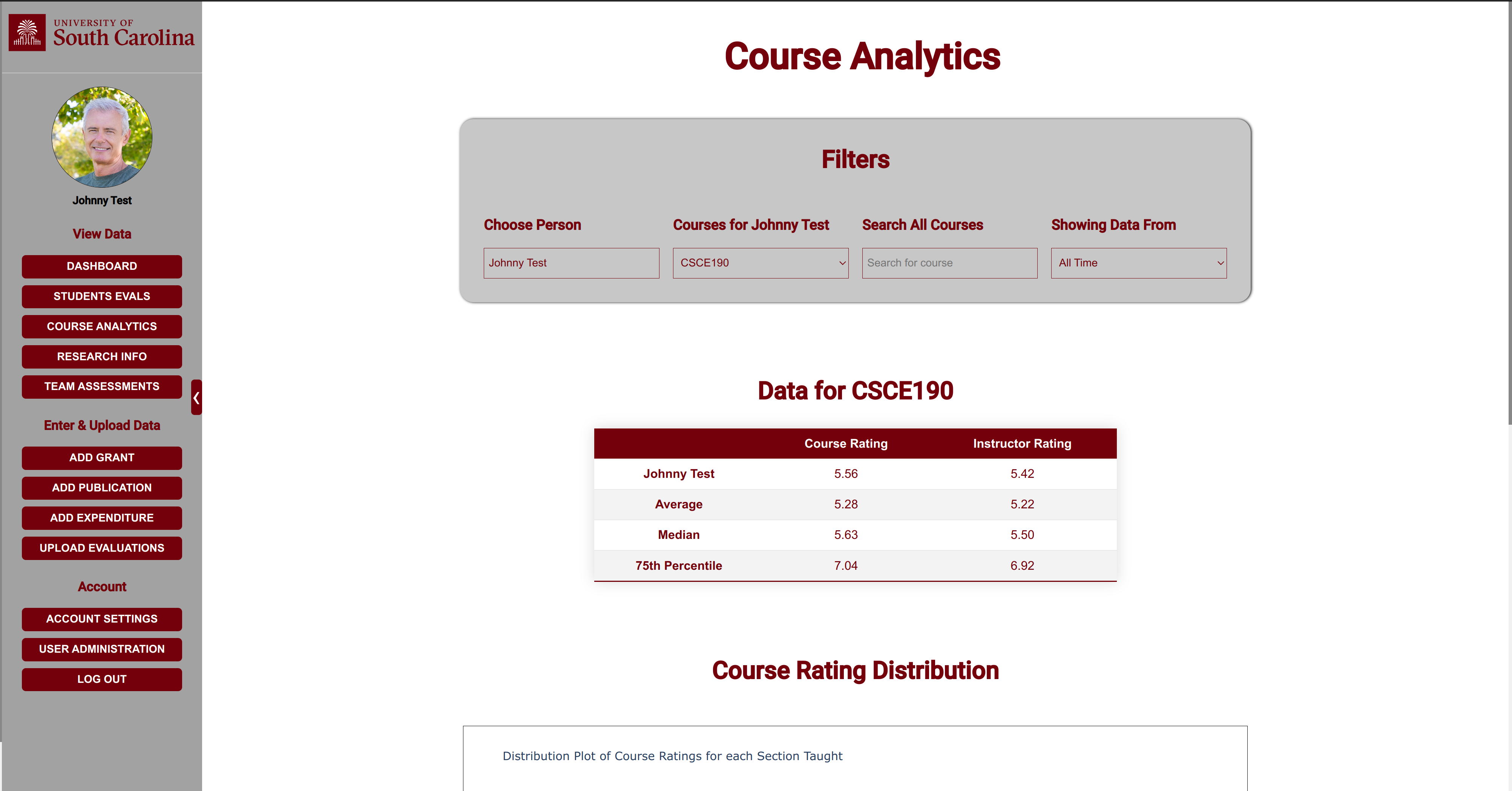}
        \end{adjustbox}
        \caption{The Course Analytics page on our platform. Users can choose a course (either one they teach or one they do not) and a time period to view results for. They are shown a table with data for the chosen course and time period, as well as--if they scroll down--anonymized distribution plots for average course ratings and average instructor ratings (Shown in Figure \ref{fig:dist_plot}). Administrators, as shown here, are given the option to view this data for members in their department.}
        \label{fig:course_analytics}
    \end{figure}

\section{Research Information}
    \begin{figure}[H]
        \centering
        \begin{adjustbox}{frame=0.25pt}
            \includegraphics[width=1\linewidth]{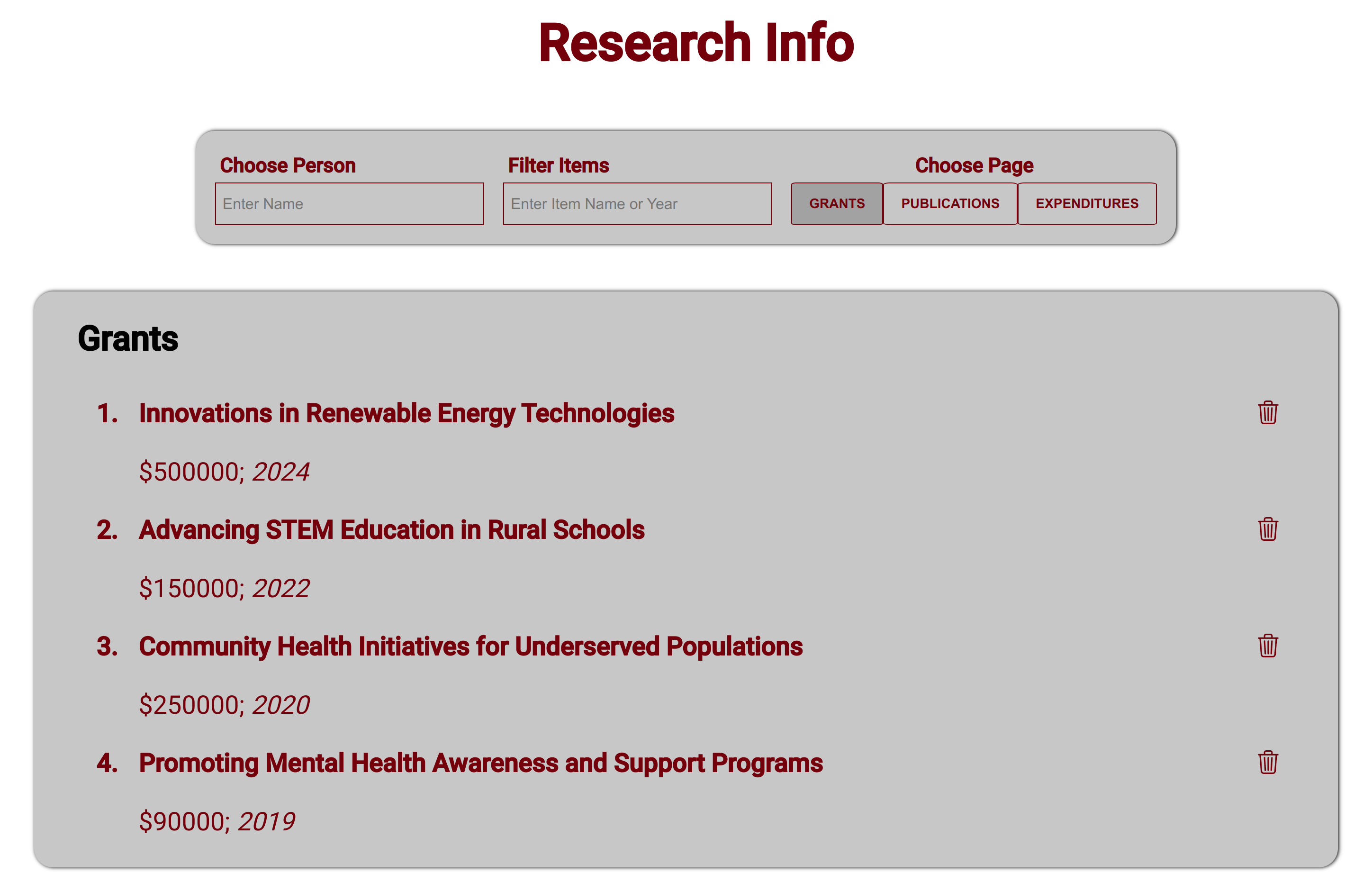}
        \end{adjustbox}
        \caption{The Research Info Page on our platform, showing the Grants tab. Similar tabs are available for publications and expenditures (a preview of which can be seen in Figure \ref{fig:dashboard}. Here, an administrator's view is shown. As such, the user is able to see their own grants or, using the ``Choose Person'' option, grants of faculty members in their department. Other faculty members would only see their own. On this page, users also have the option to filter items by a search query via the ``Filter Items'' option.}
        \label{fig:researchinfo-grants}
    \end{figure}

\section{Team Assessments}
    \begin{figure}[H]
        \centering
        \begin{adjustbox}{frame=0.25pt}
            \includegraphics[width=1\linewidth]{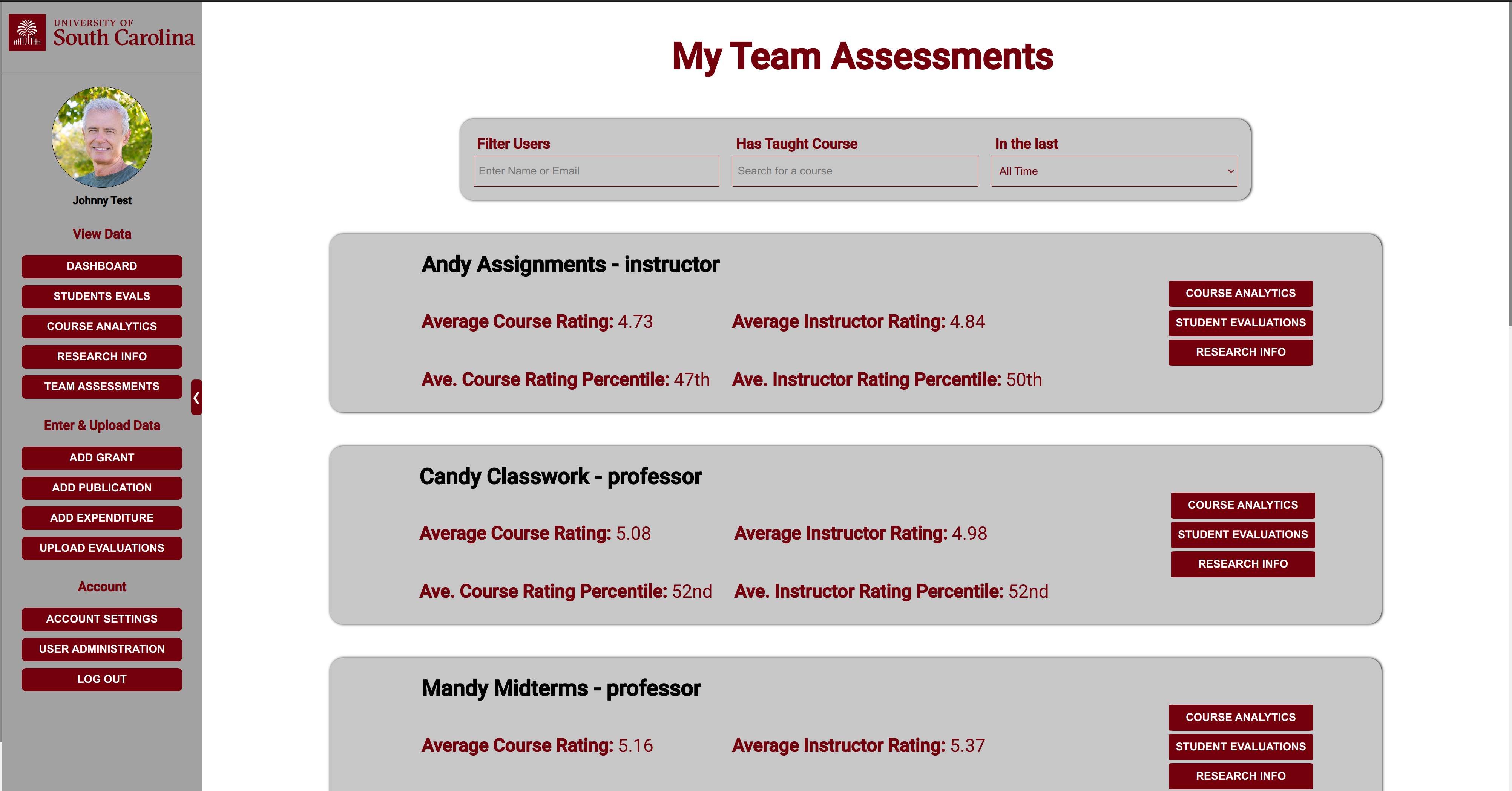}
        \end{adjustbox}
        \caption{The Team Assessments Page on our platform. Only administrators have access to this page. Here, they can view high level performance information for each member of their department. Buttons for each team member are provided to quickly navigate to other pages and view more details for that individual. Filters are available to search for members via their name or courses they teach.}
        \label{fig:team_assess}
    \end{figure}

\section{Report \& Upload Data}
    \begin{figure}[H]
        \begin{subfigure}{1\textwidth}
            \centering
            \begin{adjustbox}{frame=0.25pt}
                \includegraphics[width=0.7\linewidth]{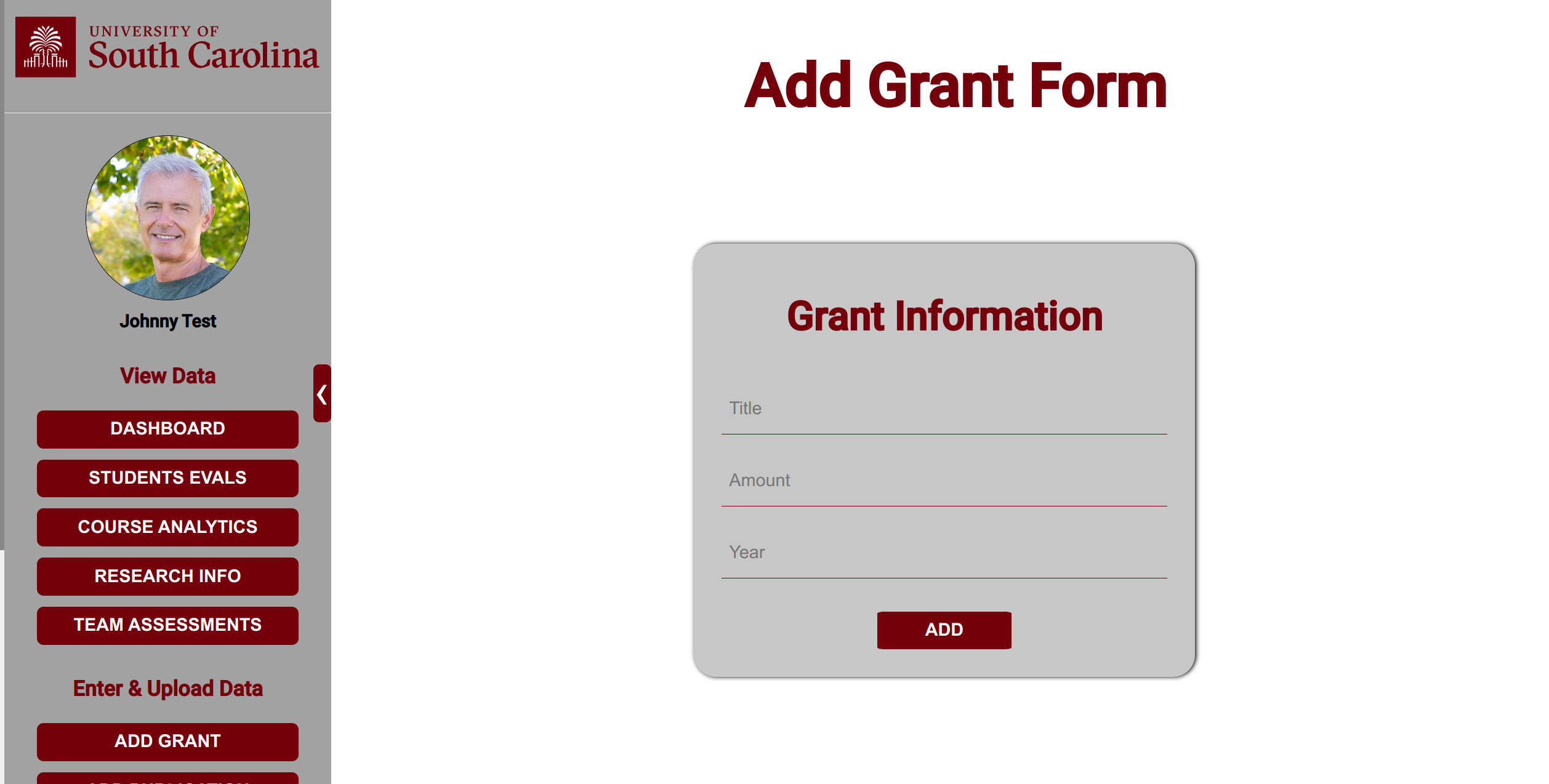}
            \end{adjustbox}
            \caption{Grants}
            \label{fig:uploadgrants}
        \end{subfigure}
        \begin{subfigure}{1\textwidth}
            \centering
            \begin{adjustbox}{frame=0.25pt}
                \includegraphics[width=0.7\linewidth]{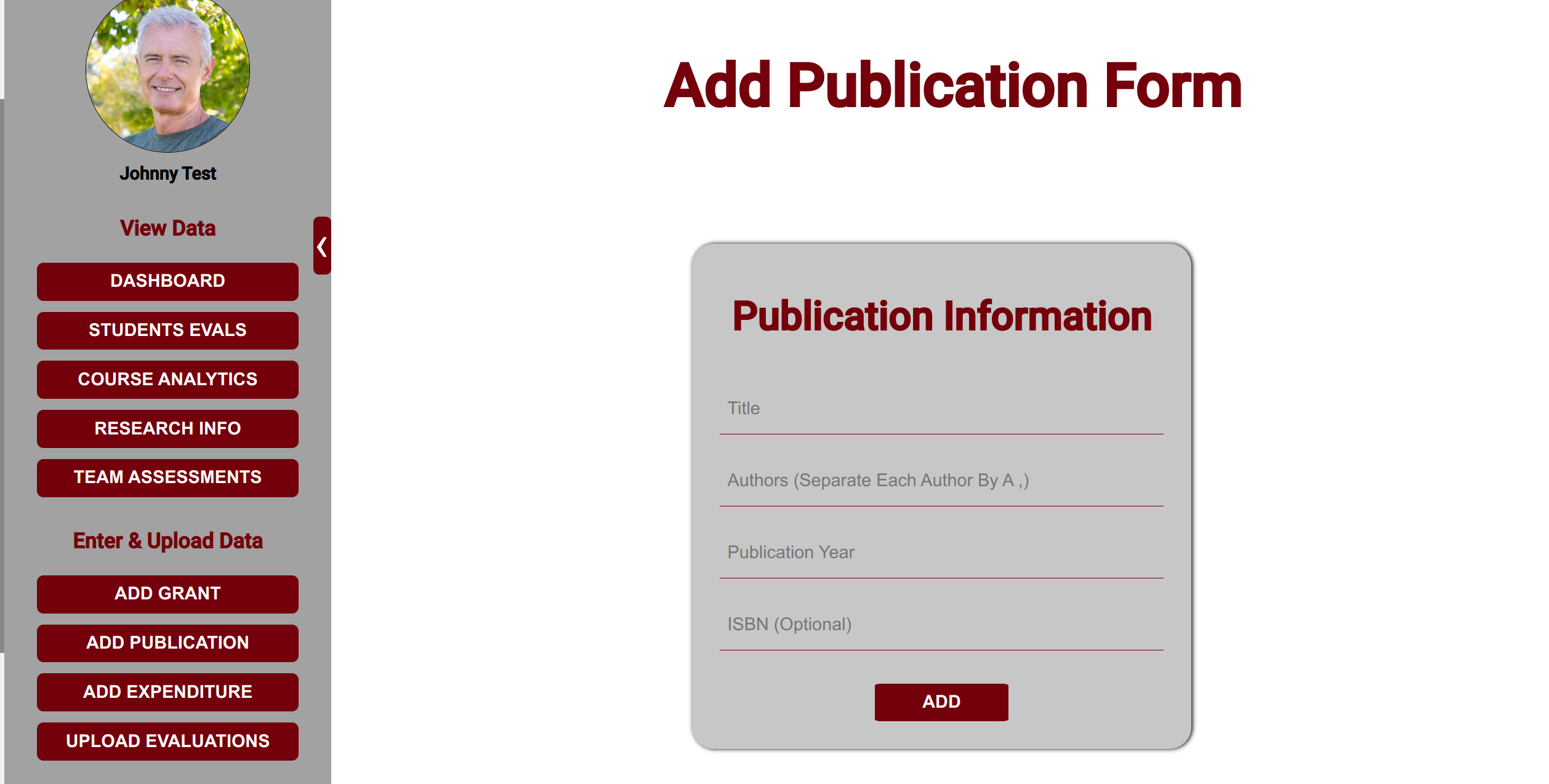}
            \end{adjustbox}
            \caption{Publications}
            \label{fig:uploadpubs}
        \end{subfigure}
        \begin{subfigure}{1\textwidth}
            \centering
            \begin{adjustbox}{frame=0.25pt}
                \includegraphics[width=0.7\linewidth]{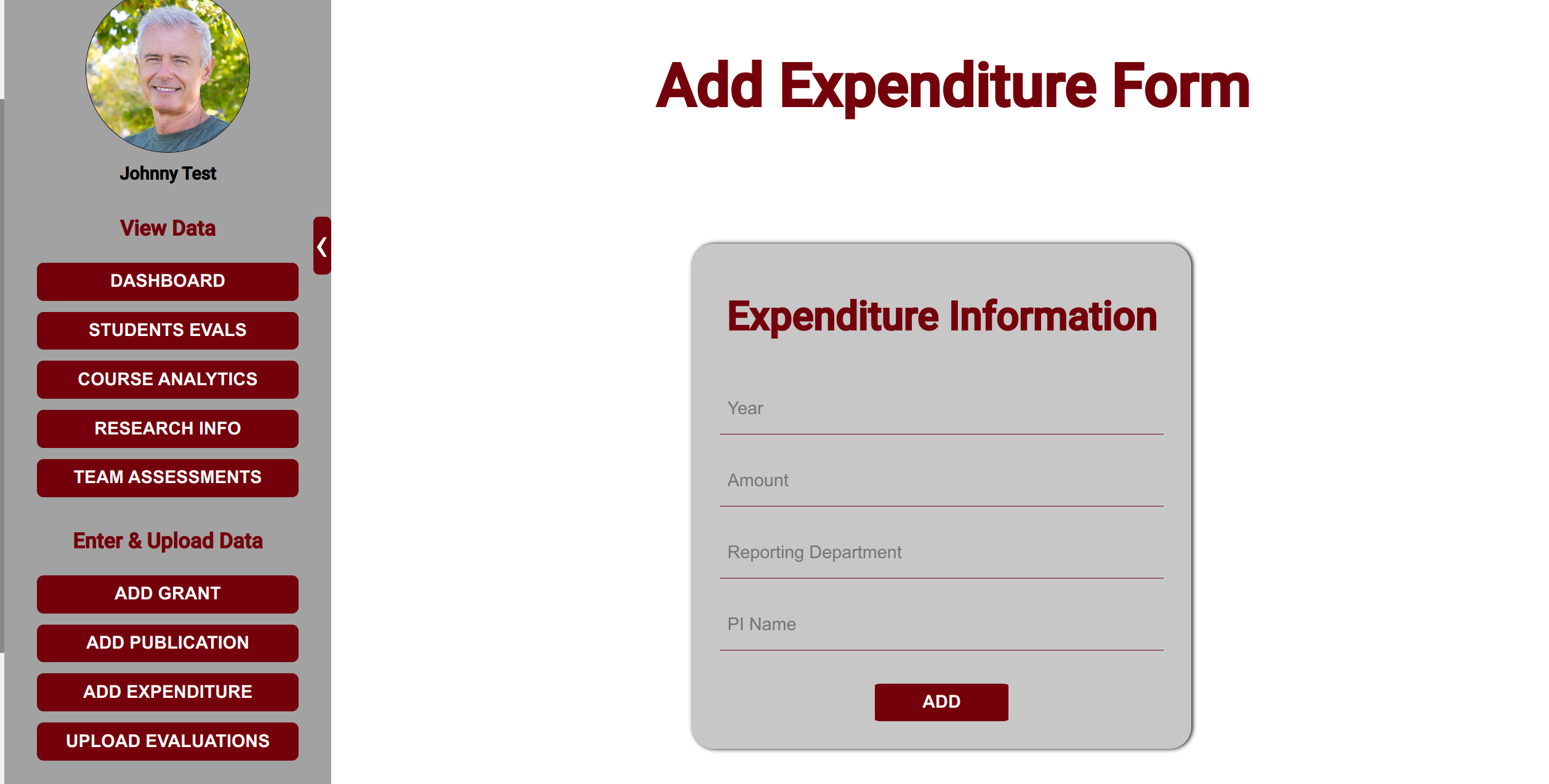}
            \end{adjustbox}
            \caption{Expenditures}
            \label{fig:uploadexpens}
        \end{subfigure}
        \caption{Pages with forms to report Grants (\ref{fig:uploadgrants}), Publications (\ref{fig:uploadpubs}), and Expenditures (\ref{fig:uploadexpens}). All users can report their information through these forms, which immediately become available for themselves and their department chair.}
        \label{fig:uploaddata}
    \end{figure}

    \begin{figure}[H]
        \centering
        \begin{adjustbox}{frame=0.25pt}
            \includegraphics[width=0.8\linewidth]{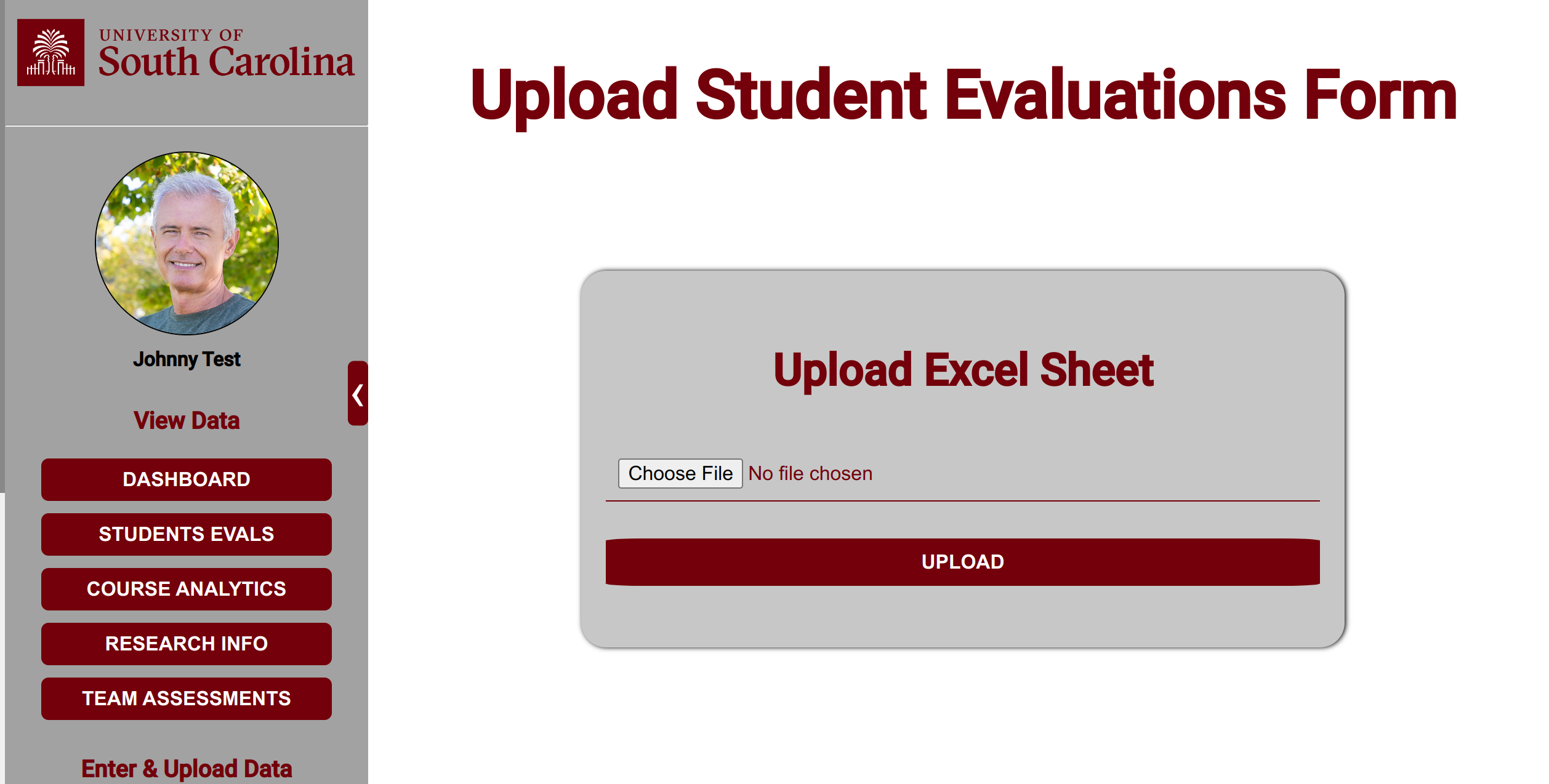}
        \end{adjustbox}
        \caption{The Upload Student Evaluations Form on our platform. Only chairs have access to this page, where they can upload the Excel sheets they receive containing student evaluations for all faculty members in their department each semester. Uploaded data immediately becomes available to all other users.}
        \label{fig:upload_evals}
    \end{figure}

\section{Account Settings \& User Administration}
    \begin{figure}[H]
        \centering
        \begin{adjustbox}{frame=0.25pt}
            \includegraphics[width=0.8\linewidth]{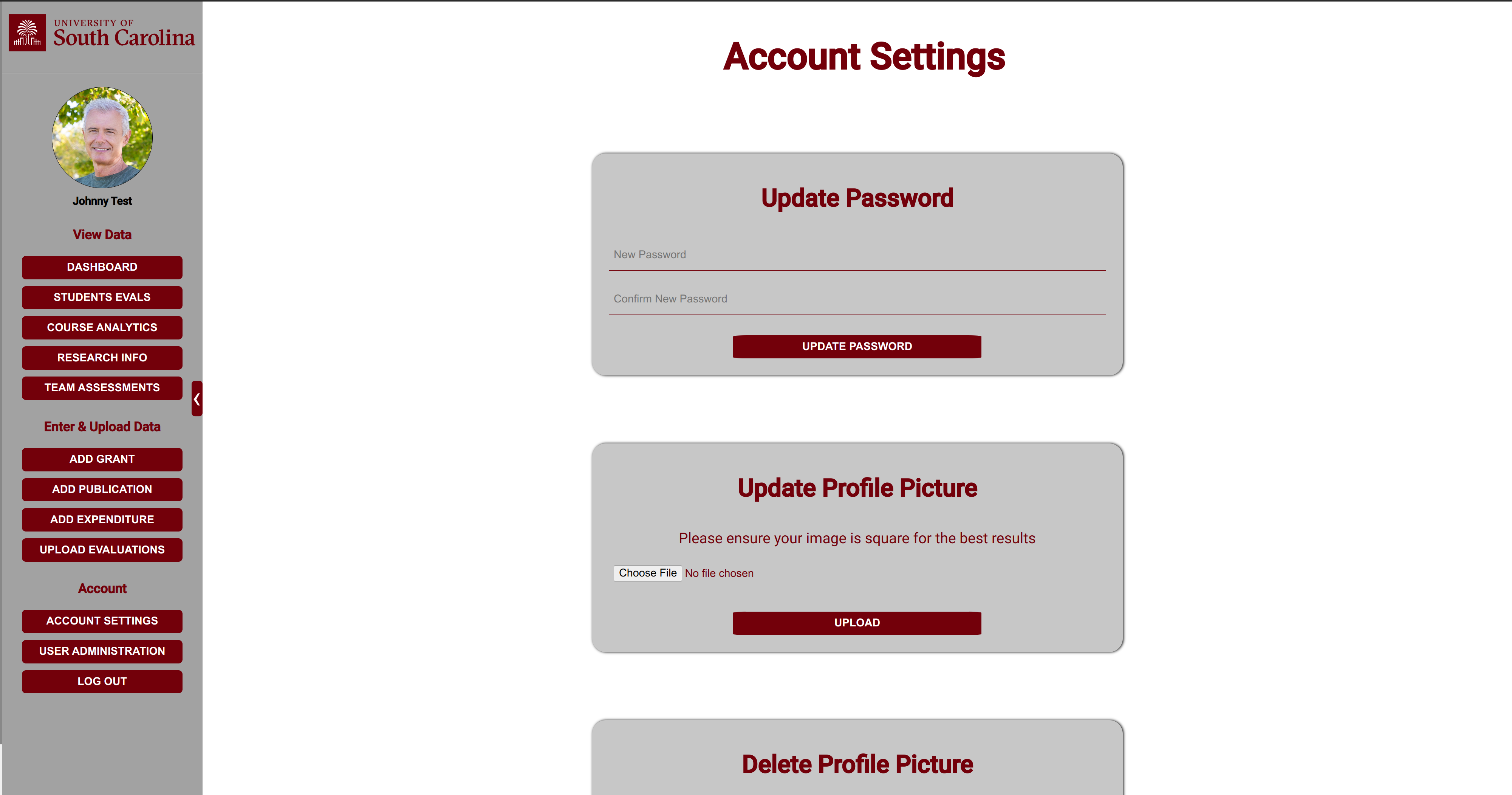}
        \end{adjustbox}
        \caption{The Account Settings page on our platform. Here, users can update their password, upload profile pictures, and delete their data.}
        \label{fig:account_settings}
    \end{figure}
    \begin{figure}[H]
        \centering
        \begin{adjustbox}{frame=0.25pt}
            \includegraphics[width=0.8\linewidth]{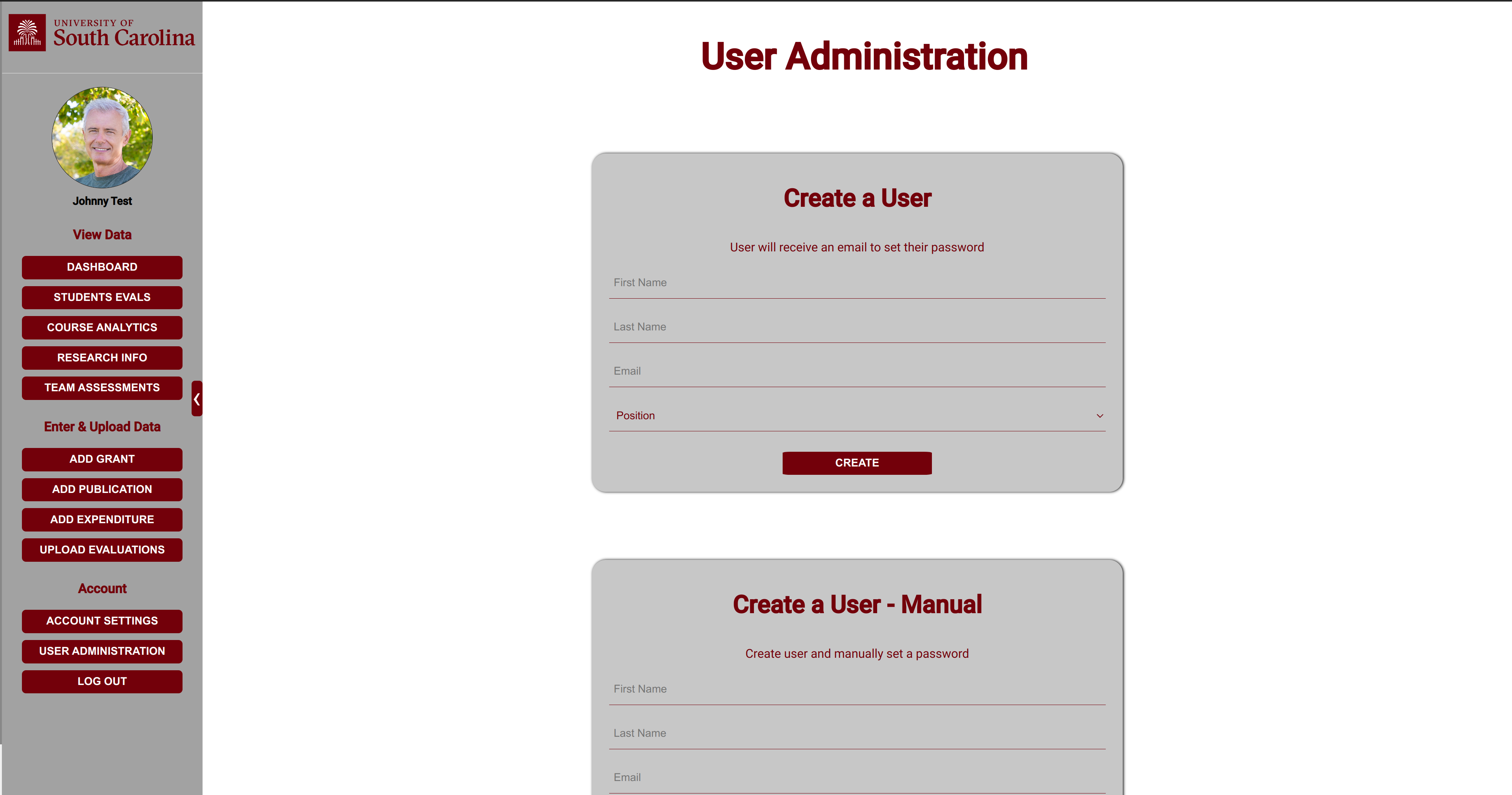}
        \end{adjustbox}
        \caption{The User Administration page on our platform. Only administrators have access to this page, where they can create, update, and delete users. Since users cannot create their own accounts, chairs have the option to create users. They can either create a user ``Manually'', where they set the password for the new account, or by email, where the new user receives an email with a link that allows them to set their own password.}
        \label{fig:user_admin}
    \end{figure}

\newpage

\end{document}